\documentclass[twocolumn,aps,floats,letterpaper,floatfix,groupedaddress]{revtex4-1}
\usepackage{graphicx}
\usepackage{dcolumn}
\usepackage{amsmath}
\usepackage{amsfonts}
\usepackage{amssymb}
\usepackage{epsfig,float,afterpage}
\usepackage{natbib}

\newcommand{\beq}{\begin{equation}}
\newcommand{\eeq}{\end{equation}}
\newcommand{\bes}{\begin{subequations}}
\newcommand{\ees}{\end{subequations}}
\newcommand{\bea}{\begin{eqnarray}}
\newcommand{\eea}{\end{eqnarray}}
\newcommand{\ba}{\begin{array}}
\newcommand{\ea}{\end{array}}
\newcommand{\beqn}{\begin{eqnarray*}}
\newcommand{\eeqn}{\end{eqnarray*}}

\newcommand{\f}[2]{\frac{#1}{#2}}

\newcommand{\om}{\omega}

\newcommand{\la}{\langle}
\newcommand{\ra}{\rangle}

\def\nn{\nonumber}

\newlength{\sizeonefig}
\newlength{\sizetwofig}
\begin{document}

\title{Cross-correlation spin noise spectroscopy of heterogeneous interacting spin systems}

\author{Dibyendu Roy$^{1,2*}$, Luyi Yang$^{3*}$, Scott A. Crooker$^3$, and Nikolai A. Sinitsyn$^1$}

\affiliation{$^1$Theoretical Division, Los Alamos National Laboratory, Los Alamos, NM 87545, USA}

\affiliation{$^2$Center for Nonlinear Studies, Los Alamos National Laboratory, Los Alamos, NM 87545, USA}

\affiliation{$^3$National High Magnetic Field Laboratory, Los Alamos National Laboratory, Los Alamos, NM 87545, USA}

\pacs{72.70.+m, 72.25.Rb, 78.67.Hc}

\begin{abstract}

We develop and apply a minimally invasive approach for characterization of inter-species spin interactions by detecting spin fluctuations alone. We consider a heterogeneous two-component spin ensemble in thermal equilibrium that interacts via binary exchange coupling, and we determine \emph{cross-correlations} between the intrinsic spin fluctuations exhibited by the two species.  Our theoretical predictions are experimentally confirmed using `two-color' optical spin noise spectroscopy on a mixture of interacting Rb and Cs alkali vapors. The results allow us to explore the rates of spin exchange and total spin relaxation under conditions of strict thermodynamic equilibrium.
\end{abstract}
\vspace{0.0cm}

\maketitle

There are numerous natural and engineered systems in which interactions between ``spins of different kind" lead to the emergence of new and interesting physics. Examples include the interaction between electron spins from different Bloch bands that gives rise to heavy-fermion behavior and Kondo-lattice effects in correlated-electron materials \cite{heavy-fermions1, heavy-fermions2}, the decoherence of solid-state spin qubits by a nuclear spin bath \cite{nuclear-bath1, nuclear-bath2, nuclear-bath3, Li12}, ferromagnetism in diluted magnetic semiconductors \cite{DMS1, DMS2}, and spin-exchange pumping of noble gas nuclei for medical imaging \cite{exchange-pumping}.

Where possible, interspecies spin interactions are generally studied by well-developed perturbative methods that, for example, selectively polarize or disturb one spin species while the influence on the other is separately monitored \cite{exchange-review}. However, interaction cross-sections often depend strongly and non-linearly on the non-equilibrium spin polarizations that are induced \cite{sensitivity}. As an alternative to conventional perturbation-based techniques for measuring spins and magnetization, methods for optical \emph{spin noise spectroscopy} \cite{Crooker04, OestreichPhysE, Zapasskii13b} have been recently developed in which electron and hole spin dynamics are revealed via the passive detection of their intrinsic and random spin fluctuations in thermal equilibrium -- \emph{i.e.}, without any polarization, excitation, or pumping.
% This noise-based approach is guaranteed by the fluctuation-dissipation theorem.

To date, spin noise spectroscopy has been applied to many different \emph{single} species of spins, such as specific alkali atoms \cite{Crooker04, Aleksandrov81, Sorensen98, Katsoprinakis07, Shah2010}, itinerant electron spins in semiconductors \cite{Oestreich05, Crooker09, Huang2011, Poltavtsev14}, and localized hole spins in quantum dot ensembles \cite{Crooker10, Li12}. These studies have shown that dynamic properties of spin ensembles such as g-factors, relaxation rates and decoherence times are measurable simply by ``listening" (typically via optical Faraday rotation) to the system's intrinsic and random spin noise, an approach ensured by the fluctuation-dissipation theorem.

Based on these developments, here we seek to explore whether spin interactions between {\it different} spin ensembles can also be directly revealed and studied -- under conditions of strict thermal equilibrium -- through their stochastic spin fluctuations alone. We envision a type of experiment shown in Fig. 1, wherein two spin species A and B in thermal equilibrium  interact, \emph{e.g.}, by spin exchange \cite{exchange2, budker-book}. If the intrinsic spin fluctuations from species A and B can be independently detected, then signatures of spin interactions may be expected to appear in the \emph{cross-correlation} of these two spin noise signals. For example, anti-correlations are expected if the interactions are purely of the spin-exchange type as depicted.

In this paper we  show that  multi-probe spin noise spectroscopy can be applied to characterize interspecies spin-spin interactions by detecting the system's intrinsic stochastic spin fluctuations.
We develop a theory for such cross-correlations in heterogeneous spin systems at thermodynamic equilibrium. In particular, we prove a universal sum rule (a ``no-go theorem")  that imposes restrictions on such cross-correlators. These results are directly compared to an experimental study of a well-understood interacting spin system (a mixture of warm Rb and Cs vapors) by applying a new type of ``two-color" spin noise spectroscopy \cite{Yang14}, and excellent agreement is found. Thus, we introduce a framework for both theoretical and experimental exploration of a broad class of heterogeneous interacting spin systems by detecting their spin fluctuations in equilibrium.

\begin{figure}
\includegraphics[width=\columnwidth]{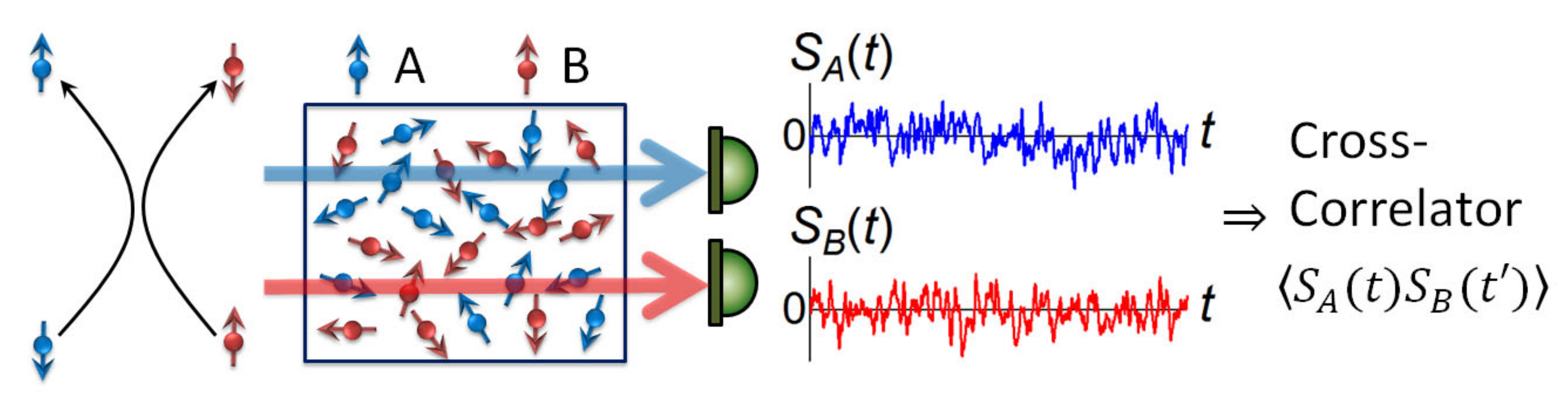}
\caption{A conceptual experiment wherein spin interactions in a heterogeneous spin system are revealed via their intrinsic spin fluctuations while in thermal equilibrium. Different probes detect spin fluctuations in the different spin species, A and B.  Interactions are revealed via cross-correlations of the two spin noise signals.}
\label{schematic}
\end{figure}

\begin{figure}[tbp]
\includegraphics[width=.45\textwidth]{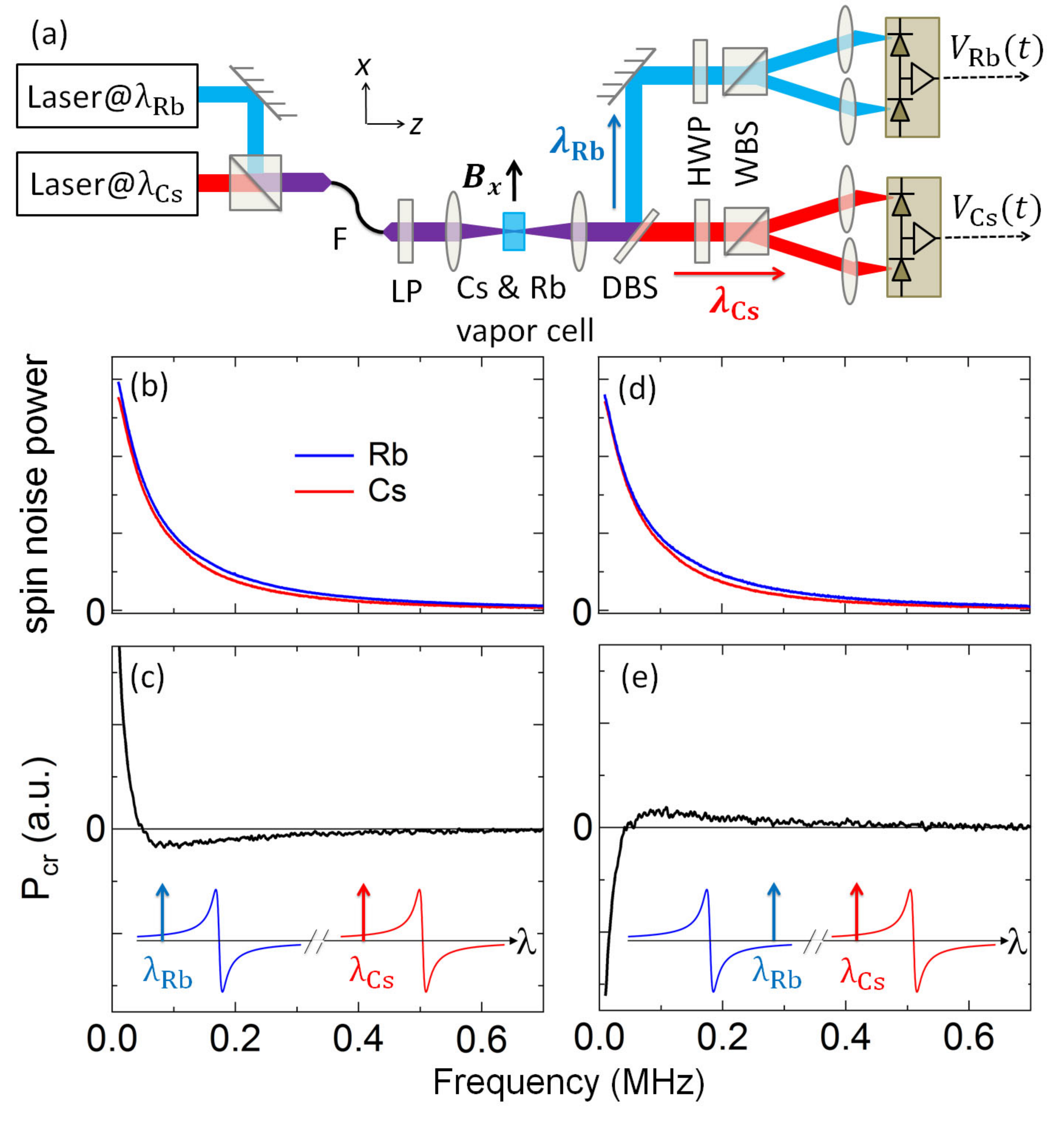}
\caption{(a) Experimental setup: two probe lasers are detuned by $\sim$100~GHz from the Rb and Cs D1 transitions (794.98~nm and 894.59~nm, respectively), then combined in a single-mode fiber (F), and focused through the Rb/Cs vapor cell. Random spin fluctuations $\delta S_z(t)$ in Rb and Cs impart Faraday rotation (FR) fluctuations $\delta \theta (t)$ on the transmitted probes, which are then separated by a dichroic beam splitter (DBS) and measured by balanced photodiodes. LP: linear polarizer, HWP: half-wave plate, WBS: Wollaston beam splitter. (b) Spin noise power density from Rb and Cs at $B_x$=0. (c) The corresponding cross-correlator $P_{\rm cr}(\om)$. (d,e) Similar, but for the case when one probe laser is detuned \emph{above} its D1 transition. Insets: cartoons of the FR induced by a polarized ground-state electron near the Rb and Cs D1 transitions, and the probe laser wavelengths $\lambda_{\rm Rb}$ and $\lambda_{\rm Cs}$.}
\label{setup}
\end{figure}

To most easily introduce the notion of spin noise spectroscopy and to describe how spin fluctuations are detected and correlated, we first describe the experiment and its results. Figure ~\ref{setup}(a) depicts the setup. A glass cell containing both Rb and Cs metal (and 100 Torr of Ar buffer gas) is heated to $\sim$140 $^\circ$C, giving a classical alkali vapor with Rb and Cs particle densities of about 0.6 and 1.4$\times10^{14}$ cm$^{-3}$.  To independently probe the intrinsic spin fluctuations in both species, we perform spin noise spectroscopy \cite{Crooker04} using \emph{two} linearly-polarized probe lasers with wavelengths $\lambda_{\rm Rb}$ and $\lambda_{\rm Cs}$ that are detuned by $\sim$100~GHz below the fundamental D1 ($^2S_{1/2}-^2P_{1/2}$) optical transitions of Rb and Cs, respectively. This large detuning significantly exceeds any Doppler or pressure broadening or hyperfine splitting of the D1 transitions, ensuring that the probe lasers do not pump or excite the atoms. Moreover, the large detuning of the probe laser and the pressure broadening of the D1 line due to the buffer gas ($\sim$10 GHz) simplifies the analysis of the data because we can ignore the hyperfine sub-structure of the D1 transition and can effectively consider the Rb and Cs atoms as having simple spin-1/2 magnetic ground states \cite{Mitsui}.

Despite the large detuning of the probe lasers, the random spin fluctuations of the Rb $5S$ and Cs $6S$ valence electrons along the $\hat{z}$ direction -- $S_{{\rm Rb},z}(t)$ and $S_{{\rm Cs},z}(t)$ -- can be detected by the optical Faraday rotation (FR) fluctuations $\theta_{\rm Rb}(t)$ and $\theta_{\rm Cs}(t)$ that they impart on the detuned probe lasers.  This detection scheme is made possible by the optical selection rules in alkali atoms, and because FR depends not on absorption but rather on the right- and left- circularly polarized indices of refraction of the alkali vapors ($\theta \propto n^R - n^L$), which decay slowly with large laser detuning \cite{Happer}. The detuned lasers can therefore be regarded as passive, non-perturbing probes of the Rb and Cs vapor's intrinsic spin fluctuations \cite{Crooker04, Kuzmich, Katsoprinakis07, Shah2010}.

The two probe lasers are combined in a single-mode optical fiber to ensure spatial overlap before being weakly focused through the vapor cell, after which they are separated by a dichroic beamsplitter. FR fluctuations $\theta_{\rm Rb}(t)$ and $\theta_{\rm Cs}(t)$ are measured by separate balanced photodiode pairs. The fluctuating output voltages $V_\alpha (t)$ ($\propto \theta_\alpha (t)$, where $\alpha$=Rb, Cs) are continuously digitized and processed in real time.  Specifically, we compute the frequency spectrum of the spin noise power density for each species $\alpha$, which is equivalent to the Fourier transform of the spin-spin correlator:
 \bea
P_{\alpha}(\omega)=\int_{-\infty}^{\infty}dt~e^{i\omega t} \la S_{\alpha}(t)S_{\alpha}(0) \ra.
\label{SNPS-A}
\eea

Importantly, we also compute the real part of the cross-correlation spectrum between the Rb and Cs spin fluctuations:
\begin{equation}
%\begin{split}
P_{\rm cr}(\om)=\int_{-\infty}^{\infty}dt\, e^{i\omega t} \left[ \la S_{{\rm Rb}}(t)S_{{\rm Cs}}(0) \ra + \la S_{{\rm Cs}}(t)S_{{\rm Rb}}(0)\ra \right],
\label{CSNPS}
%\end{split}
\end{equation}
which has not been considered previously for spin noise studies but which, as shown below, contains specific information about inter-species spin coupling and interactions. Note that for clarity, the subscript `$z$' was omitted from all the spin projections $S_z(t)$ in Eqs. 1 and 2.

Finally, small static magnetic fields $B_x$ can be applied along the transverse ($\hat{x}$) direction, which forces the spin fluctuations $S_z(t)$ to precess, thereby shifting the measured spin noise to higher (Larmor) frequencies.

Figure ~\ref{setup}(b) shows the power spectra of the detected spin noise from the Rb and Cs spins at $B_x$=0 [$P_{\rm Rb}(\omega)$ and $P_{\rm Cs}(\omega)$; blue and red curves respectively]. These spin noise peaks are centered at zero frequency and they exhibit approximately Lorentzian lineshapes. Their linewidths are dominated here by the effective transit-time broadening of the atoms across the diameter of the focused probe lasers in the vapor. The intensity and detuning of the probe beams were adjusted here to give approximately equal Rb and Cs spin noise power.

Figure ~\ref{setup}(c) shows the corresponding and simultaneously-measured noise cross-correlation spectrum between the two species, $P_{\rm cr}(\omega)$.  Crucially, $P_{\rm cr}(\omega)$ is \emph{not} zero, indicating that interspecies spin interactions do appear in and are measurable through intrinsic spin fluctuations alone. $P_{\rm cr}(\omega)$ exhibits a very narrow peak centered at zero frequency, revealing \emph{positive} correlations between Rb and Cs spin fluctuations at small frequencies. In addition, $P_{\rm cr}(\omega)$ also exhibits a broader negative feature at larger frequencies, revealing \emph{anti}-correlations between Rb and Cs spins at these frequencies.  Importantly, $P_{\rm cr}(\omega)$ can be fit extremely well by the difference of two Lorentzians with equal area (\emph{i.e.}, $P_{\rm cr}$ has zero total integrated area), the origin and significance of which is discussed below.

To confirm these cross-correlation signals, Figs. \ref{setup} (d,e) show similar measurements acquired when one of the probe laser wavelengths is tuned \emph{above} (rather than below) its corresponding D1 transition. While the noise power spectra for the individual vapors are unaffected as expected, $P_{\rm cr}(\omega)$ inverts sign because the FR induced by a polarized ground-state alkali spin is an odd function of wavelength about the D1 transition (see inset diagrams). Moreover, it was verified that $P_{\rm cr}(\omega)$=0 when the two probe beams were spatially separated in the vapor (not shown).

\begin{figure}[tbp]
\includegraphics[width=.45\textwidth]{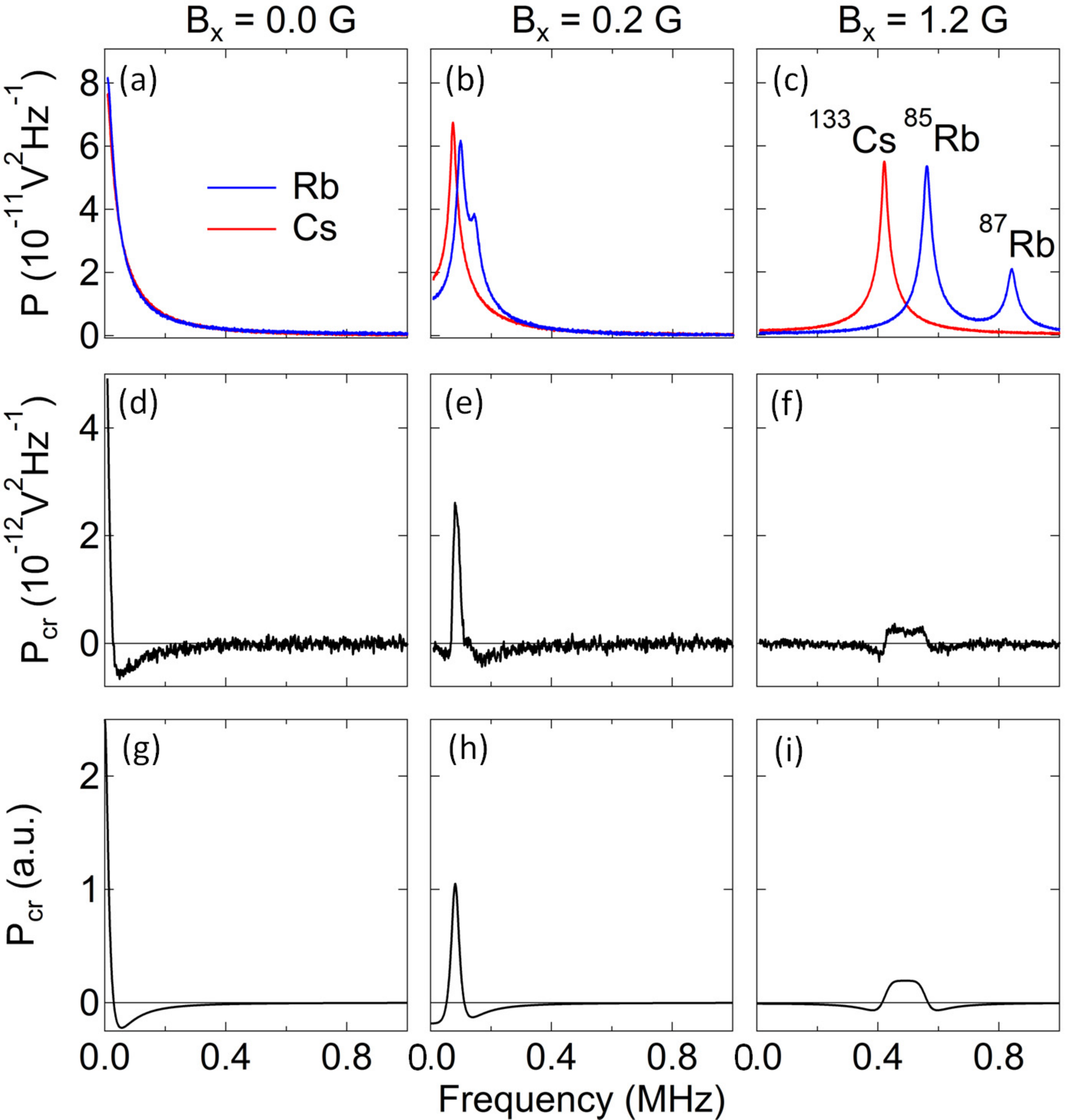}
\caption{(a-c) Measured spin noise power spectra $P_{\rm Rb}(\omega)$ and $P_{\rm Cs}(\omega)$ at $B_x$=0, 0.2, 1.2 G; (d-f) corresponding cross-correlation spectra $P_{\rm cr}(\omega)$. (g-i) Calculation of $P_{\rm cr}(\omega)$ [from Eq.~(\ref{CSNPS3})], using $N_A=N_B$ and spin flip rates $\gamma_A$=$\gamma_B\equiv\gamma_1$=15 kHz, and $\gamma_2$=60 kHz.}
\label{field}
\end{figure}

Figures 3(a-c) show the measured spin noise power spectra from Rb and Cs at different values of $B_x$. With increasing $B_x$, the noise peaks shift to higher frequency (due to precession) at different rates in accord with their $g$-factors (1/3 and 1/4, respectively). At 0.2~G the Rb and Cs spin noise peaks are still largely overlapped, while at 1.2~G they are mostly separated.  A higher-frequency spin noise peak from the less abundant $^{87}$Rb isotope ($g$=1/2) is also visible \cite{Crooker04}. Importantly, Figs. 3(d-f) show that the corresponding cross-correlator $P_{\rm cr}(\omega)$ also shifts to higher frequencies, diminishes in amplitude, and develops a more complex structure.  At larger $B_x$ when the Rb and Cs spin noise peaks no longer overlap at all, $P_{\rm cr}(\omega)$ disappears entirely (not shown).  Finally, Figs. 3(g-i) show $P_{\rm cr}(\omega)$ calculated from the theoretical model that is developed immediately below.

In order to model and understand these experiments, we develop and apply a theory for interpreting cross-correlations between spin fluctuations in a two-component spin-$\frac{1}{2}$ ensemble. We introduce vectors ${\bf S}_A$ and ${\bf S}_B$ whose components are the total (unnormalized) spin polarization along the $x,y,z$-axes of type A and B spin in the observation volume. Given numbers $N_{Az\uparrow}$ and $N_{Az\downarrow}$ of $\uparrow$ and $\downarrow$ spins of type A, then $S_{Az}=(N_{Az\uparrow}-N_{Az\downarrow})/2$. We now formulate a useful sum rule, which is valid irrespective of further model details:

\underline{No-Go Theorem}: {\it At thermodynamic equilibrium and in the limit of large spin temperature, the integral of the cross-correlator over frequency is zero.}
This theorem shows that useful information about interactions between system components is contained only in the functional form of $P_{\rm cr}(\om)$. In the large temperature limit, it rules out strategies based only on measurements of integrated noise power (\emph{e.g.}, \cite{Dellis13}).

 \underline{Proof}: The integral of $P_{\rm cr}(\om)$ over $\om$, defined in (\ref{CSNPS}), gives a delta function in time, which is removed by integration over time to produce a cross-correlator at $t$=0:
\begin{equation}
\int_{-\infty}^{\infty} P_{\rm  cr}(\omega) d\, \omega = 2\pi \la \{ S_{Az}(t),S_{Bz}(t) \}_{t=0} \ra,
\end{equation}
where curly brackets are the anti-commutator. %that emerges when spin polarizations are treated as quantum mechanical operators
%If the temperature scale significantly exceeds all possible spin coupling energies, configurations of spin densities with opposite signs (independently for each species) are equally probable. Hence the thermal average of equal time spin correlators of different species are zero in this limit. In other words,
The equilibrium spin density matrix at large temperature is proportional to a unit matrix. The trace of its product with a traceless operator, such as $S_{Az}S_{Bz}$, is zero.
{Q.E.D.}

To model spin interactions and the essential role of spin fluctuations, we first assume that species A and B each have an intrinsic spin relaxation process with rates $\gamma_{A}$ and $\gamma_{B}$ per particle (due to, \emph{e.g.}, interactions with cell walls, buffer gas, etc).
%In a monoatomic vapor of A-atoms,  %co-flips do not change to total spin polarization, and hence  are not observable.
In a short time interval $dt$, the average change of the spin polarization, \emph{e.g.} $\la \delta S_{Az}(t)\ra $, due to such processes is  $\la \delta S_{Az}(t)\ra=-\gamma_{A} S_{Az}(t) dt$.
Since such processes are independent for different atoms, we have $\la (\delta S_{Az}(t))^2\ra - \la \delta S_{Az}(t)\ra^2 =\gamma_{A} N_{A}dt$. A fluctuation with such a variance can be produced by
a white noise source $\eta_{Az}(t)$, with a correlator $\langle \eta_{Az}(t) \eta_{Az}(t')\rangle = \gamma_{A} N_{A} \delta(t-t')$. The assumption of Gaussian white noise is justified here because $N_{A,B} \gg 1$.
%For simplicity, we assume the case of a weak magnetic field in comparison to hyperfine coupling.
 %Combining these facts,  the dynamics of the polarization of a monoatomic vapor is described by a stochastic equation
%\bea
%\f{d {\bf S}_{A}}{dt}=-\gamma_A{\bf S}_{A}+{\boldsymbol \eta}_{A},
%\label{SDP1}
%\eea
%with $\la\eta_{Ai}(t)\ra=0$ and $\la \eta_{Ai}(t)\eta_{Aj}(t')\ra=\gamma_AN_{A}\delta_{ij} \delta(t-t')$, $i,j=x,y,z$.
%We can write similar stochastic dynamical equations for the transverse spin-polarizations $P_{Ay}(t),P_{Bx}(t)$ and $P_{By}(t)$.
%
%

Next, spin-exchange interactions between A and B spins lead to total-spin-conserving {\it co-flip processes} with a rate $\gamma_{AB}$ per each pair of particles of different kinds.
%Such scatterings conserve the total spin but transfer spins between A and B, leading to cross-correlations of their spin fluctuations.
 The average change in $S_{Az}(t)$ during short time interval $dt$ due to co-flip processes is given by $\la \delta S_{Az}(t)\ra=-\gamma_{AB} dt [S_{Az}(t)N_B-S_{Bz}(t)N_A]/2$, where the $1/2$ appears because, on average, only half of the opposite species participates in the exchange interaction with a given atom.
The variance of such fluctuations is $\la (\delta S_{Az}(t))^2\ra -\la  \delta S_{Az}(t)\ra^2 =\gamma_{AB}N_AN_Bdt/2$ , which can be represented by another white noise source $\eta_{AB,z}(t)$.
Importantly, $\la \delta S_{Az}(t) \delta S_{Bz}(t)\ra-\la \delta S_{Az}(t)\ra \la \delta S_{Bz}(t)\ra=-\gamma_{AB}N_AN_Bdt/2$.
%The latter correlator is negative, reflecting the fact that random co-flips create fluctuations of different  sign in different species.
% between different species is important to understand the dynamics of transverse polarizations in the presence of inter species spin-exchange collisions. Generally $P_{Ax}P_{Bx}\ll N_AN_B$, therefore we write $\la (\delta P_{Ax}(t))^2\ra=\gamma_{AB}N_AN_B/2$ and $\la \delta P_{Ax}(t)\delta P_{Bx}(t)\ra=-\gamma_{AB}N_AN_B/2$.

Combining the intrinsic spin dynamics with random inter-species co-flip processes, we find \cite{note2}:
\bea
\f{d{\bf S}_{A}}{dt}&=&g_A{\bf S}_A\times {\bf B}-\gamma_A{\bf S}_{A}-\f{\gamma_{AB}}{2}(N_B{\bf S}_A-N_A{\bf S}_B)\nn\\&+&{\boldsymbol \eta}_{A}+{\boldsymbol \eta}_{AB},\label{SDP2}\\
\f{d{\bf S}_{B}}{dt}&=&g_B{\bf S}_B\times {\bf B}-\gamma_B{\bf S}_{B}-\f{\gamma_{AB}}{2}(N_A{\bf S}_B-N_B{\bf S}_A)\nn\\&+&{\boldsymbol \eta}_{B}-{\boldsymbol \eta}_{AB},\label{SDP3}
\eea
where $g_{A,B}$ are the gyromagnetic ratios of species A and B. Equations 4 and 5 are standard rate equations but with the crucial and explicit inclusion of fluctuation terms ${\boldsymbol \eta}$. These noise sources are correlated as
\bea
\noindent \la \eta_{\alpha i}(t)\eta_{\beta j}(t')\ra&=&\delta_{\alpha \beta}\delta_{ij}N_\alpha\gamma_{\alpha}\delta(t-t'), \label{noiseprop1}\\
\la \eta_{AB,i}(t)\eta_{AB,j}(t')\ra&=&\delta_{ij}\delta(t-t')\gamma_{AB}N_AN_B/2,
% -\la \eta_{AB,i}(t)\eta_{BA,j}(t')\ra &=&  \delta_{ij}\delta(t-t')\f{\gamma_{AB}N_AN_B}{2}, \label{noiseprop2}
\eea
where $i,j=x,y,z$ and $\alpha,\beta=A,B$. We assume that only transverse magnetic fields $B_x$ are applied.

We define ${\bf S}(\om)\equiv{ \lim \limits_{T_m \rightarrow \infty}} (1/\sqrt{T_m})\int_{0}^{T_m} dt\, e^{i\om t}{\bf S}(t)$, where $T_m$ is the measurement time. By taking the Fourier transform of Eqs.~(\ref{SDP2}),~(\ref{SDP3}), and averaging over noise, we obtain
% determine the spin correlators, we note that Eq.~(\ref{Seq}) corresponds to a Wannier process \cite{gardener} whose
spin correlators at the steady state:  %%correlator for components of ${\bf P}$
\begin{equation}
\langle (S_{\alpha i} (\omega) S _{\beta j} (-\omega) \rangle = \left( \frac{1}{{\bf R}-i\omega{\bf 1}} {\bf G}   \frac{1}{{\bf R}^{\rm {\bf T}}+i\omega {\bf 1}} \right)_{ \alpha i,\beta j},
\label{gard}
\end{equation}
%with $i,j$ are running through the set $x, y,z$ for each valley.
where $i,j=x, y,z$;  $\alpha, \beta=A,B$; ${\bf 1}$ is a unit matrix.
Introducing %notation $\gamma'_B=\gamma_{AB}N_B/2,~\gamma'_A=\gamma_{AB}N_A/2$, and
 a bar operation, $\bar{A}=B$ and $\bar{B}=A$, and the Levi-Civita symbol $\varepsilon_{ijk}$, matrices ${\bf R}$ and ${\bf G}$ then read:
\bea
R_{\alpha \beta }^{ij} &=& \delta_{ij} \left[ \delta_{\alpha \beta} \gamma_{\alpha} +\frac{\gamma_{AB}(\delta_{\alpha \beta}N_{\bar{\beta}}  -\delta_{\alpha\bar{\beta}} N_{\alpha} )}{2} \right]- \delta_{\alpha \beta} g_\alpha B_x  \varepsilon_{xij},\nn\\
 G_{\alpha \beta }^{ij}&=& \delta_{ij} \left[ \delta_{\alpha \beta} \gamma_{\alpha} N_\alpha  +\frac{\gamma_{AB}N_A N_B}{2} ( \delta_{\alpha \beta} - \delta_{\alpha \bar{\beta}}) \right].
\nn
\eea
The cross-correlator (\ref{CSNPS}) is then given by: %) and  the individual spin noise powers (\ref{SNPS-A})
\begin{equation}
\noindent P_{\rm cr}(\om)=\la S_{Az}(\om) S_{Bz}(-\om) \ra+[A \Leftrightarrow B] ,
\label{CSNPS2}
\end{equation}  % \la S_{Bx}(\om)S_{Ax}(-\om)\ra
% \begin{equation}
% \noindent P_{\alpha} (\om)= \la |S_{\alpha z}(\om)|^2 \ra, \quad \alpha=A,B.
%\label{s-total}
%\end{equation}
%\la (\alpha \tilde{P}_{Ax}(\om)+\beta \tilde{P}_{Bx}(\om))(\alpha \tilde{P}_{Ax}(-\om)+\beta \tilde{P}_{Bx}(-\om))\ra\nn \\
%\begin{figure}
%\includegraphics[width=\columnwidth]{theory.eps}
%\caption{Numerical prediction of Eq.~(\ref{gard}) for the form of the cross-correlator, $P_{\rm cr}$,  of an atomic mixture at three values of the  magnetic field: $B_x=0$, $0.2$ and $1.2$ Gauss. The ratio of atoms in the mixture is chosen as $N_A:N_B=1:1$, and the respective gyromagnetic ratios are $g_A=$0.35 MHz/G and $g_B=$0.47 MHz G$^{-1}$. The spin flip rates are assumed to be $\gamma_1=0.015$ MHz, and $\gamma_2=0.06$ MHz.}%$\gamma_A:\gamma_B=\gamma'_A:\gamma'_B=3:1$, and $\gamma_B=\gamma'_B=0.167$ kHz. The coupling strength $\alpha=\beta=1$.}
%\label{SN1}
%\end{figure}
%Full expressions for the cross-correlator (\ref{CSNPS2})   can be written explicitly  but it would be bulky and not illuminating. Instead, we will explore it numerically and then
%discuss specific limits for which a greatly simplified description can be derived.
A compact expression can be obtained by assuming identical values for relaxation rates: $\gamma_A=\gamma_B \equiv \gamma_1$:
%, where we introduced parameter $\gamma$ so that it can be directly compared with the parameter $\gamma_{AB}$.
\bea
P_{\rm cr}(\om)=N_AN_B \gamma_{AB}  \sum \limits_{s=\pm} \f{\chi_{s} }{\chi_{s}^2+\kappa_{s}^2},   %+\f{\chi_-}{\chi_-^2+\kappa_-^2}\big),
\label{CSNPS3}
\eea
where $\chi_{\pm}=2[\gamma_1\gamma_2-(\om \pm \Omega_A)(\om \pm \Omega_B)],~\kappa_{\pm}=2(\om\pm\Omega_A)\Gamma_B+2(\om\pm\Omega_B)\Gamma_A$; $\Omega_{A,B}\equiv g_{A,B}B_x$
% and $\Omega_B \equiv g_BH$
are the Larmor frequencies of the spin species and where
%We also define the broadening due to the intrinsic and co-flip rates,
\begin{equation}
%\gamma_1&=&\gamma(N_A+N_B), \quad
\gamma_2=\gamma_1+\gamma_{AB}\frac{N_A+N_B}{2},\,\,\,
\Gamma_{A,B}=\gamma_1+\gamma_{AB}\frac{N_{B,A}}{2}.
\label{times}
\end{equation}
Figures~\ref{field}(g-i) show $P_{\rm cr}(\om)$ calculated according to Eq.~(\ref{CSNPS3}). Although the model does not include the additional $^{87}$Rb isotope, it shows good agreement with experimental data in Fig.~\ref{field}(d-f). We now discuss three different limits of Eq.~(\ref{CSNPS3}).
\begin{figure}[tbp]
\includegraphics[width=.45\textwidth]{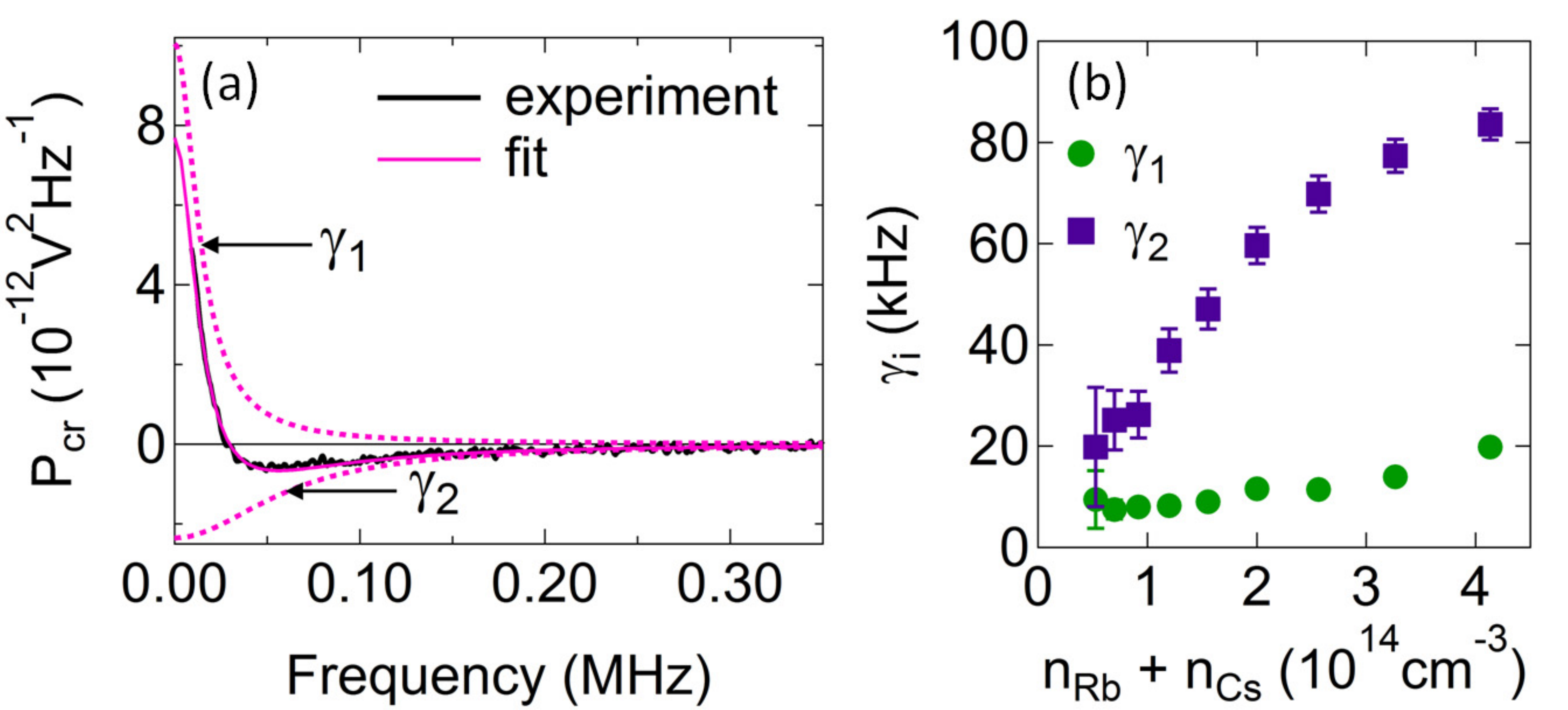}
\caption{$P_{\rm cr}(\omega)$ measured at $B_x$=0, fit with two Lorentzians of equal and opposite area (dashed lines), in agreement with the `no-go' theorem. (b) Relaxation rates $\gamma_{1,2}$ extracted from the fit by Eq.~\ref{CSNPS4} versus the total vapor density $n_{\rm Rb}+n_{\rm Cs}$. Approximately linear dependence is in agreement with the assumption of pairwise spin interactions.}
\label{exp2}
\end{figure}

% $\tau_A^{-1}=(N_A\gamma_{AB}/2+1/\tau),~\tau_B^{-1}=(N_B\gamma_{AB}/2+1/\tau),~\tau^{-1}=\gamma(N_A+N_B)$ and $\tau_0^{-1}=(\gamma+\gamma_{AB}/2)(N_A+N_B)$.
(i) First, at a zero applied magnetic field, we find % the { cross-correlator  spectrum} is given by
\begin{equation}
P_{\rm  cr} (\om)=%\f{2N_AN_B}{N_A+N_B}
2 Q \Big(\f{\gamma_1}{\om^2+\gamma_1^2}-\f{\gamma_2}{\om^2+\gamma_2^2}\Big),
 \label{CSNPS4}
\end{equation}
which is simply the difference of two equal-area Lorentzians with widths $\gamma_{1}$ and $\gamma_{2}$. Here $Q= N_AN_B/(N_A+N_B)$.
%in agreement with Fig.~\ref{field}(d,g)
%This result is consistent with the shape of Fig.~\ref{SN1}(d) for $P_{\rm \text corr}(\om)$ obtained for low values of the external field. %It is straightforward to check now that the total area under $P_{\rm \text corr}^{\Omega} (\om)$ is zero.
%A straightforward consequence of (\ref{CSNPS4}) is that if $\gamma_{2} \gg \gamma_1$, i.e. if the exchange co-flips are much faster than the spin-nonconserving scatterings, then  the negative part of $P_{\rm \rm cross}(\om)$ is much broader and  much shallower than the positive part.

In Fig.~\ref{exp2}(a) we show the experimentally measured $P_{\rm cr}(\omega)$ at $B_x$=0. In good agreement with Eq.~\ref{CSNPS4} and the `no-go' theorem, $P_{\rm cr}(\omega)$ has zero total area, being well fit by the difference of two equal-area Lorentzians (dashed lines). Figure ~\ref{exp2}(b) shows the extracted $\gamma_{1,2}$ as a function of the total vapor density $n_{\rm Rb}+n_{\rm Cs}$, which is tuned with the cell temperature.
Here we recall that $\gamma_1$ describes the relaxation of the total spin $S_{Az} + S_{Bz}$. The spin exchange rate is therefore characterized by the difference $\gamma_2-\gamma_1$. As is typical for alkali vapors, we find that the relaxation rate of the total spin is much smaller than the exchange rate, since the latter conserves the total spin. Now we can interpret the negative part of $P_{\rm cr}(\omega)$ (\ref{CSNPS4}) as emerging from the expected anti-correlations induced by fast spin co-flips between Rb and Cs atoms.  On the other hand, the positive-valued peak in (\ref{CSNPS4}) and in the data is due to the fact that, at fast co-flip rate, the total spin polarization is equally observed by both beams at longer time scales (\emph{i.e.}, the total spin relaxation is `shared' between the interacting Rb and Cs atoms), which corresponds to positive cross-correlations.

 (ii) Next, we consider the limit $\gamma_1 \ll (g_A-g_B)B_x \ll \gamma_{2}$,
 which is close to the case measured in Figs.~\ref{field}(b,e).  Here, one can disregard effects of $\gamma_1$ and obtain
\begin{equation}
P_{\rm cr}(\om)=Q \
\sum_{s=\pm 1}\Big[\f{\gamma_1^{\prime}}{(\om-s\Omega_L)^2+\gamma_1^{\prime 2}}-\f{\gamma_2}{(\om-s\Omega_L)^2+\gamma_2^2}\Big],
\label{CSNPS5}
\end{equation}
where
\begin{equation}
\Omega_L=\frac{\Omega_AN_A+\Omega_BN_B}{N_A+N_B}, \,\, \gamma_1^{\prime}=\frac{(\Omega_A-\Omega_B)^2N_AN_B}{\gamma_{2}(N_A+N_B)^2},
\label{avom}
\end{equation}
which indicates that the spectrum shifts to an effective (weighted average) Larmor frequency $\Omega_L$, while the positive-valued peak is broadened by the magnetic field (an effective total spin relaxation rate), in agreement with Figs.~\ref{field}(e,h).

(iii) In the limit of a large magnetic field, Eq.~(\ref{CSNPS3}) predicts that the cross-correlator $P_{\rm cr}(\om)$ vanishes, also in agreement with experimental observation.

Thus, we see that the theoretical model is confirmed by the experimental data and therefore captures the essential physics of spin fluctuations and their correlations.  We note, however, that a more rigorous and quantitative description of the observed noise power and cross-correlations should include not only the presence of all different isotopes but also the coupled dynamics of their nuclear and electronic spins -- \emph{i.e.}, the fact that alkali atoms actually have a nontrivial magnetic ground state due to hyperfine splitting -- which would lead to multiple correlated resonances even within the same atomic species. On the other hand, the experimental results apparently validate our simple two-component approximation. This can be explained, for example, by the presence of fast intra-species spin-exchange interactions that smear the physics related to presence of multiple resonances from the same atomic species. We also note that, in order to achieve a better theoretical precision, it is straightforward to amend our approach to include multiple intra-species resonances by extending the set of equations (\ref{SDP2})-(\ref{SDP3}).

In summary, we have shown, both experimentally and theoretically, that cross-correlations between the stochastic spin fluctuations of different species reveals specific information about spin interactions. Crucially, these interactions can be detected using unperturbed spin ensembles under conditions of strict thermal equilibrium. Such non-invasive characterization techniques may find future applications in metrology, \emph{e.g.} to reveal the physics that limits the efficiency of various magnetometers \cite{Shah2010, budker-book}. We also envision applications of this technique to mixtures of ultra-cold atomic gases and condensates \cite{Dalfovo99, Ni08}, which are sensitive to the probe interference \cite{Meineke12}.
Studies of cross-correlations of spin noise in solid state physics, \emph{e.g.} in multiple Bloch bands and in new layered materials as well as in artificial semiconductor nanostructures, represent additional and as-yet-unexplored avenues for applications of cross-correlation studies and two-color spin noise spectroscopies.

We gratefully acknowledge helpful discussions with Igor Savukov and Darryl Smith, and support from the Los Alamos LDRD Program.


\begin{thebibliography}{34}
\bibitem{note2} We note that a related model was explored in \cite{Dellis13} but for  different  noise sources in  Eqs.~(\ref{SDP2})-(\ref{SDP3}) and in a context different from two-color spin noise spectroscopy.
\end{thebibliography}
\end{document}